\documentclass[aps,prl,amsmath,amssymb,amsfonts,nofootinbib,long]{revtex4}
\usepackage{epsfig,latexsym,bm,makeidx}

\begin{document}

\title{ANISOTROPIC DARK ENERGY AND ELLIPSOIDAL UNIVERSE}

\author{L. Campanelli$^{1}$}
\email{leonardo.campanelli@ba.infn.it}

\author{P. Cea$^{1,2}$}
\email{paolo.cea@ba.infn.it}

\author{G. L. Fogli$^{1,2}$}
\email{gianluigi.fogli@ba.infn.it}

\author{L. Tedesco$^{1,2}$}
\email{luigi.tedesco@ba.infn.it}

\affiliation{$^1$Dipartimento di Fisica, Universit\`{a} di Bari, I-70126 Bari, Italy}
\affiliation{$^2$INFN - Sezione di Bari, I-70126 Bari, Italy}

\date{January, 2011}


\begin{abstract}
\noindent
{\bf Abstract.} A cosmological model with anisotropic dark energy is analyzed. The amount of deviation from
isotropy of the equation of state of dark energy, the skewness $\delta$, generates an
anisotropization of the large-scale geometry of the Universe, quantifiable by means of the
actual shear $\Sigma_0$. Requiring that the level of cosmic anisotropization at the time of
decoupling is such to solve the ``quadrupole problem'' of cosmic microwave background
radiation, we find that $|\delta| \sim 10^{-4}$ and $|\Sigma_0| \sim 10^{-5}$, compatible
with existing limits derived from the magnitude-redshift data on type Ia supernovae.
\end{abstract}


\maketitle


\section{\normalsize{I. Introduction}}
\renewcommand{\thesection}{\arabic{section}}

The isotropy of the cosmic microwave background (CMB) radiation, first
seen by the Cosmic Background Explorer (COBE) satellite~\cite{COBE1} and then reinforced by the
Wilkinson Microwave Anisotropy Probe (WMAP) data~\cite{WMAP1}, together with the assumption that we are
not in any special position in the Universe, underlies the Cosmological Principle,
according to which we live in a homogeneous and isotropic universe described by a
{\it Robertson-Walker line element}.

Tiny deviations from perfect isotropy, at the level of $10^{-5}$, have been also reported by
COBE~\cite{COBE2} and thereafter confirmed by the high resolution WMAP data. The observed CMB anisotropy
spectrum is in impressive agreement with the predictions of the $\Lambda$CDM concordance model,
namely the widely accepted standard model of (inflationary) cosmology~\cite{Weinberg}.

Nevertheless, data from 7-yr WMAP observations~\cite{WMAP7a} do not have removed an effect seen at lower
significance in the COBE data, that is a lack of power on the CMB anisotropy quadrupole moment.

Since measurements at the largest angular scales are affected by foreground and other
systematic effects, we have to wait for the full results of the PLANCK mission~\cite{PLANCK}
that will check, using a different frequency coverage and scanning strategy than WMAP,
the anomalously low quadrupole so far observed.

Surprisingly, however, analyzing the first light sky map released by the PLANCK team,
which covers about only the $10\%$ of the sky, Liu and Li~\cite{Liu}
have recently claimed that the amplitude of CMB power spectrum on large scales is
significantly lower than that reported by the WMAP team.

The ``smallness'' of CMB quadrupole amplitude, referred to as {\it quadrupole problem},
signals a suppression of power at large scales and has been intensively studied~\cite{Quadrupole}
meanly because it may signal either a nontrivial topology or a deviation from isotropy
of the large scale geometry of
the Universe. Indeed, recently enough, it has been shown that a
spatially homogeneous but anisotropic cosmological model of Bianchi type I
described by a plane-symmetric line element, the {\it ellipsoidal universe}~\cite{prl,prd},
allows a better matching of the large-scale CMB anisotropy data.

Various mechanisms could have induced a planar symmetry in the spatial geometry of the universe:
topological defects (e.g. cosmic stings, domain walls)~\cite{Berera},
a uniform cosmological magnetic field~\cite{prl,prd},
magnetic fields possessing planar symmetry at cosmic scales~\cite{prd2}, or a
moving~\cite{Beltran Jimenez} dark energy.

In particular, the mechanism proposed by Koivisto and Mota of dark energy
with anisotropic equation of state~\cite{Koivisto-Mota,Koivisto-Mota2}
(see also Refs.~\cite{Barrow,Rodrigues,Akarsu,Appleby,Battye}) is very attractive,
since cosmic anisotropy originates from the actual dominant component of the
Universe and then could be directly tested, for example, by either observations of
magnitude and redshift of type Ia supernovae or cosmic parallax effects 
of distant sources~\cite{Quercellini}.

The aim of this paper is to connect the dark energy anisotropization mechanism to the
ellipsoidal universe proposal. As we will see, the deviation from isotropy of the equation
of state of dark energy triggers an anisotropization of the Universe
which can be well described by an ellipsoidal universe model.
The level of such an anisotropization can be then chosen to solve the quadrupole problem.

The plan of the paper is as follows. In section II, we briefly review the
ellipsoidal universe proposal. In section III, we introduce a cosmological
model with an anisotropic dark energy component and we study its connection
with both the ellipsoidal universe model and the quadrupole problem.
Finally, we draw our conclusions in section IV.


\section{\normalsize{II. Ellipsoidal Universe: CMB Quadrupole}}
\renewcommand{\thesection}{\arabic{section}}

Planar symmetry, in a cosmological context, is described by the {\it Taub line element}~\cite{Taub}:
\begin{equation}
\label{metric}
ds^2 = dt^2 - a^2(t) (dx^2 + dy^2) - b^2(t) \, dz^2,
\end{equation}
where the scale factors $a$ and $b$ are normalized as $a(t_0) =
b(t_0) = 1$ at the present (cosmic) time $t_0$.

As shown in Ref.~\cite{prl}, the above metric generates a quadrupole term
in the CMB radiation, through the Sachs-Wolfe effect, which adds to that
caused by the inflation-produced gravitational potential.

More precisely, let us introduce the CMB temperature anisotropy
in terms of spherical harmonics $Y_{lm}$,
\begin{equation}
\label{DeltaT} \Delta T({\bf n}) =
\sum_{l=2}^{\infty} \sum_{m=-l}^{l} a_{lm} Y_{lm}({\bf n}),
\end{equation}
where ${\bf n}$ is a unit direction vector. The coefficients of the expansion,
$a_{lm}$, are related to the angular power spectrum, $C_l$, through
\begin{equation}
\label{Cl} C_l = \frac{1}{2l + 1} \sum_{m=-l}^{l} |a_{lm}|^2,
\end{equation}
and the quadrupole ($l=2$) component is defined as
\begin{equation}
\label{QC2} \mathcal{Q} \, \equiv \, \sqrt{\frac{3}{\pi} \, C_2} \, .
\end{equation}
The observed CMB anisotropy map is then a linear superposition of two
contributions~\cite{Bunn}:
\footnote{As pointed out in Ref.~\cite{Koivisto-Mota2},
this is an approximation which is good when the spacetime background anisotropy is small
(this will be indeed our case, see section III), and that
a rigorous treatment of CMB anisotropies should presuppose the analysis of
perturbations in the anisotropic background Eq.~(\ref{metric}).}
\begin{equation}
\label{alm} a_{lm} = a_{lm}^{\rm A} + \, a^{\rm I}_{lm},
\end{equation}
where $a_{lm}^{\rm A}$ comes from the anisotropic spacetime background,
while the $a^{\rm I}_{lm}$ term is the standard isotropic fluctuation due to
inflation-produced gravitational potential at the last scattering surface.

The 7-year WMAP data give, for the quadrupole amplitude, the value~\cite{WMAP7a}
\begin{equation}
\label{QWMAP}
\mathcal{Q}_{\rm obs}^2 \simeq 200 \, \mu \mbox{K}^2,
\end{equation}
while their best-fit for the isotropic $\Lambda$CDM standard cosmological model
gives~\cite{WMAP7a}:
\begin{equation}
\label{QIWMAP} \mathcal{Q}_{\rm I}^2 \simeq 1200 \, \mu \mbox{K}^2.
\end{equation}
Even taking into account the cosmic variance~\cite{Durrer},
which dominates theoretical uncertainties at low $l$,
\begin{equation}
\label{CosmicVariance}
\sigma_{\rm cosmic} = \sqrt{\frac25} \: \mathcal{Q}_{\rm I}^2 \simeq 759 \, \mu \mbox{K}^2,
\end{equation}
the quadrupole amplitude remains anomalously low.

The quadrupole anisotropy associated to the anisotropic spacetime background
has been instead calculated in Ref.~\cite{prl}:
\begin{equation}
\label{QA} \mathcal{Q}_{\rm A} = \frac{2}{5 \sqrt{3}} \: T_{\rm cmb} e_{\rm dec}^2 \, ,
\end{equation}
where $T_{\rm cmb} \simeq 2.73$ K is the actual (average)
CMB temperature and $e_{\rm dec} = e(t_{\rm dec})$ is the ``eccentricity''
\begin{equation}
\label{ecc-def} e =
    \left\{ \begin{array}{ll}
            \sqrt{1-(b/a)^2}, &  a > b,
            \\
            \sqrt{1-(a/b)^2}, &  a < b,
    \end{array}
    \right.
\end{equation}
evaluated at the time of decoupling $t_{\rm dec}$.

The ``ellipsoidal universe'' proposal could indeed explains
the ``low'' value of the observed quadrupole, Eq.~(\ref{QWMAP}), compared to that predicted by
the standard cosmological model, Eq.~(\ref{QIWMAP}). This, however,
requires a suitable orientation of the anisotropy associated to planar geometry with respect to
the inflation-produced one, in such a way to lower the overall power to a sufficient extent.
As shown in Refs.~\cite{prd,prd2} this is attained for eccentricities approximatively given by
\begin{equation}
\label{ecc-dec} e_{\rm dec}^2 \simeq \frac{\sqrt{15} \,
[ \, 3\sqrt{73} - 5 \, \mbox{sgn} (a-b) \, ]}{24} \, \frac{\mathcal{Q}_{\rm I}}{T_{\rm cmb}} \, ,
\end{equation}
where $\mbox{sgn} \, x$ is the sign function ($\mbox{sgn} \, x = \pm 1$ if $x\gtrless 0$).
In the Appendix we show that the probability that such an orientation occurs by chance is, indeed,
not negligible.

In the next Section, we show that an anisotropic dark energy component
causes an anisotropization of the Universe described by the line element~(\ref{metric}).
In particular, the skewness, which parameterizes the
deviation from isotropy of the equation of state of dark energy,
and the shear, which quantify the level of cosmic anisotropization, will be connected
to the amount of eccentricity at decoupling.


\section{\normalsize{III. Anisotropic Dark Energy: Skewness and Shear}}
\renewcommand{\thesection}{\arabic{section}}

The most general energy-momentum tensor consistent with planar
symmetry is
\begin{equation}
\label{tensor}
T^{\mu}_{\,\,\, \nu} = \mbox{diag} \, (\rho,-p^{\|},-p^{\|},-p^{\bot}),
\end{equation}
where $\rho$ is the energy density, while $p^{\|}$ and $p^{\bot}$ are
the ``longitudinal'' and ``normal'' pressures.
Taking into account the energy-momentum tensor~(\ref{tensor}), the
Einstein's equations read
\begin{eqnarray}
\label{E1} && 3(1-\Sigma^2) H^2 = 8\pi G \rho, \\
\label{E2} && 3(1-\Sigma + \Sigma^2) H^2 + [(2-\Sigma) H]^{\cdot} = -8\pi G p^{\|}, \\
\label{E3} && 3(1+\Sigma)^2 H^2 + 2[(1+\Sigma) H]^{\cdot} = -8\pi G p^{\bot},
\end{eqnarray}
where a dot denotes a differentiation with respect to the cosmic time. Here,
we have introduced, in the usual way, the cosmic shear, $\Sigma$,
and the ``mean Hubble parameter'', $H$, as
\begin{equation}
\label{ShearHubble} \Sigma \equiv (H_a - H)/H, \;\;\;
H \equiv \dot{A}/A,
\end{equation}
where $H_a \equiv \dot{a}/a$ and $A \equiv (a^2b)^{1/3}$ is the
``mean expansion parameter''.

Instead of considering Eqs.~(\ref{E3}) and (\ref{E2}), we will analyze
the equation obtained by subtracting side by side them,
\begin{equation}
\label{EqShear} 3 (H\Sigma)^{\cdot} + 9H^2\Sigma =
8\pi G (p^{\|} - p^{\bot})
\end{equation}
(from which it is apparent that the source of the shear is
proportional to the difference between the longitudinal and
normal pressures of the anisotropic fluid), and the equation
coming from the conservation of the energy-momentum tensor,
$T^{\mu}_{\,\,\, \nu \, ;\mu} = 0$, which gives
\begin{equation}
\label{EqEnergy} \dot{\rho} + 3H \! \left( \rho + \frac{2p^{\|} + p^{\bot}}{3} \right)
+ 2H (p^{\|} - p^{\bot}) \Sigma = 0.
\end{equation}
To proceed further, we assume that the anisotropic fluid defined by
the energy-momentum tensor~(\ref{tensor}) is indeed made up of three different components:
an isotropic radiation component ($r$), an isotropic dark matter component ($m$),
and an anisotropic dark energy (DE) component,
\begin{eqnarray}
\label{rho} && \rho = \rho_r + \rho_m + \rho_{\rm DE}, \\
\label{parallelo} && p^{\|} = p_r + p_m + p^{\|}_{\rm DE}, \\
\label{perpendicolare} && p^{\bot} = p_r + p_m + p^{\bot}_{\rm DE},
\end{eqnarray}
with equations of state: $p_{r} = \rho_{r}/3$, $p_{m} = 0$, and
\begin{equation}
\label{EoSDE} p^{\|}_{\rm DE} \equiv w^{\|} \rho_{\rm DE},
\;\;\; p^{\bot}_{\rm DE} \equiv w^{\! \bot} \rho_{\rm DE}.
\end{equation}
Moreover, we will assume that the $w^{\|}$ and $w^{\!\bot}$ coefficients
are constants and that the three components are noninteracting.
The latter assumption ensures that each component is separately conserved,
so that Eq.~(\ref{EqEnergy}) gives
\begin{eqnarray}
&& \rho_r = \rho_r^{(0)} A^{-4}, \;\;\; \rho_m = \rho_m^{(0)} A^{-3}, \\
\label{EqDE} && \dot{\rho}_{\rm DE} + \left[ 3(1+w) + 2\delta \Sigma \right] H \rho_{\rm DE} = 0,
\end{eqnarray}
where from now on an index ``0'' defines quantities evaluated at the actual time,
and we have introduced the ``mean equation of state parameter'' $w$ and ``skewness'' $\delta$
as
\begin{equation}
\label{wdelta} w \equiv (2w^{\|} + w^{\!\bot})/3, \;\;\; \delta \equiv w^{\|} - w^{\!\bot}.
\end{equation}
%
%
Finally, we assume that $\Sigma$ and $\delta$ are small quantities
(as we will verify {\it a posteriori}) so that we can neglect the second term in the
square brackets of Eq.~(\ref{EqDE}). In this case, Eq.~(\ref{EqDE}) simply gives:
\begin{equation}
\label{EqDE2} \rho_{\rm DE} = \rho_{\rm DE}^{(0)} A^{-3(1+w)}.
\end{equation}
Hereafter, we use this approximate result for the evolution of dark energy density.

Introducing the ``mean density parameters''
\begin{equation}
\label{OmegaX} \Omega_X \equiv \rho_X^{(0)}/\rho_c^{(0)}, \;\;\; \rho_c^{(0)} \equiv \frac{3H_0^2}{8\pi G} \, ,
\end{equation}
where $X=r,m,\mbox{DE}$, and taking into account Eqs.~(\ref{rho})-(\ref{OmegaX}),
the shear equation~(\ref{EqShear}) gives
\begin{equation}
\label{Sigmat}  \Sigma(A) = \frac{\Sigma_0 + (E-E_0) \, \delta}{A^3H/H_0} \, ,
\end{equation}
where
\begin{equation}
\label{EqH} H(A)/H_0 = \sqrt{\Omega_r A^{-4} + \Omega_m A^{-3} + \Omega_{\rm DE} A^{-3(1+w)}}
\end{equation}
and we have defined the function
\begin{equation}
\label{xi} E(A) = \Omega_{\rm DE} \! \int_0^A \!\! \frac{dx}{x^{1+3w} H(x)/H_0} \, .
\end{equation}
It is worth noting that the density parameters are not all independent, sice evaluating
Eq.~(\ref{EqH}) at the present time gives
\begin{equation}
\label{Omegatot} \Omega_r + \Omega_m + \Omega_{\rm DE} = 1.
\end{equation}
In order to not to spoil the predictions of the standard isotropic model (such as
those coming from Big Bang Nucleosynthesis and Large Scale Structure formation),
we only consider anisotropic cosmological models which isotropize at early times.
Therefore, we impose the isotropization condition:
\begin{equation}
\label{Isotropization} \lim_{A \rightarrow 0} \Sigma(A) = 0.
\end{equation}
Taking into account Eqs.~(\ref{Sigmat})-(\ref{xi}), it is easy to verify
that the above condition is satisfied if and only if
\begin{equation}
\label{Sigma0}  \Sigma_0 = E_0 \delta,
\end{equation}
where we have taken into account that the dark energy component is subdominant
with respect to the radiation one for $A \rightarrow 0$. For completeness, we
give the asymptotic expansion of the shear for small values of the expansion parameter:
\begin{equation}
\label{SigmaAsmall} A \ll 1: \;\;
\Sigma(A) \simeq \frac{\delta}{2-3w} \, \frac{\Omega_{\rm DE}}{\Omega_r} \, A^{1-3w}.
\end{equation}
Moreover, since dark energy dominates over the other components in the far future,
$A \rightarrow \infty$, we get from Eq.~(\ref{Sigmat}):
\begin{equation}
\label{SigmaInfty}  \Sigma_\infty \equiv \lim_{A \rightarrow \infty} \Sigma(A) = \frac{2 \, \delta}{3(1-w)} \, ,
\end{equation}
where we used Eq.~(\ref{Sigma0}).

Finally, taking into account Eqs.~(\ref{Sigmat})-(\ref{xi}) and Eqs.~(\ref{Sigma0})-(\ref{SigmaInfty}),
it is possible to show that the following inequality holds:
\begin{equation}
\label{Sigma<delta}  \max_{A \in (0,\infty)} |\Sigma(A)| = |\Sigma_\infty|.
\end{equation}
Equations~(\ref{Sigma0})-(\ref{Sigma<delta}) makes evident the fact that an anisotropy
in the equation of state of dark energy ($\delta$) generates an anisotropy in the
cosmic geometry ($\Sigma$), and that the Universe will never isotropize,
although its level of anisotropization is very low being $\delta$ a small quantity.

In Fig.~1, we plot the shear as a function of the expansion parameter for various
values of $w$ and $\Omega_{\rm DE}$. Here and in the following we take
$\Omega_r = 0.83 \times 10^{-4}$~\cite{Kolb}.


\begin{figure}[t!]
\begin{center}
\includegraphics[clip,width=0.7\textwidth]{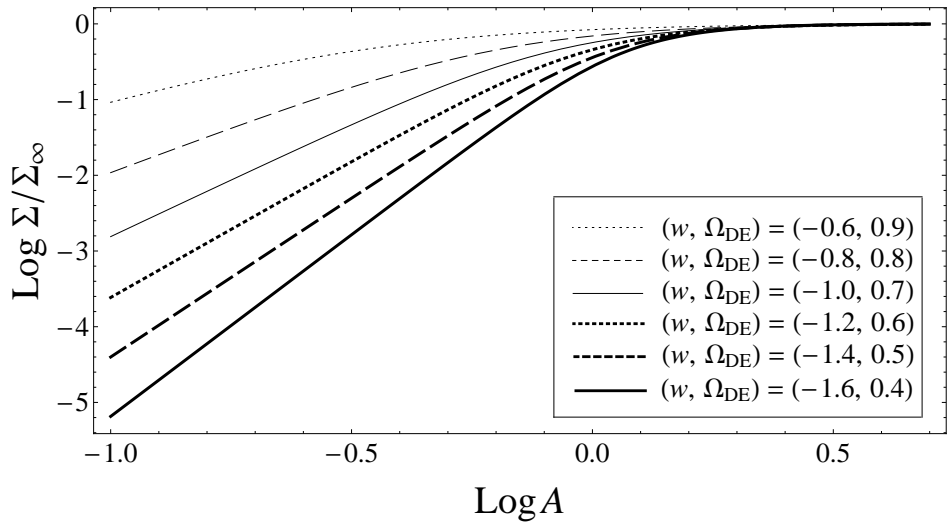}
\caption{The shear $\Sigma$ normalized to its maximum value $\Sigma_\infty$
as a function of the mean expansion parameter $A$ for different values of the
mean dark energy equation of state parameter $w$ and the mean dark energy
density $\Omega_{\rm DE}$.}
\end{center}
\end{figure}


To proceed further, we note that the eccentricity and the shear
are connected by the following relation:
\begin{equation}
\label{EccSigma} e^2 = 6 \, \mbox{sgn} (a-b)
\int_1^A \frac{dx}{x} \, \Sigma(x),
\end{equation}
valid for small eccentricities, $e \ll 1$, and coming from definitions~(\ref{ecc-def})
and (\ref{ShearHubble}). The above equation calculated at the time of decoupling,
together with Eqs.~(\ref{Sigmat}) and (\ref{Sigma0}), implies that
$\mbox{sgn} (a-b) = -\mbox{sgn} \, \delta$ and allow us to write $\delta$ and
$\Sigma_0$ in the form:
\begin{eqnarray}
\label{delta-ecc} |\delta| \!\!& = &\!\! c_\delta(w,\Omega_{\rm DE}) \: e_{\rm dec}^2,
\\
\label{Sigma0-ecc} |\Sigma_0| \!\!& = &\!\! c_\Sigma(w,\Omega_{\rm DE}) \: e_{\rm dec}^2,
\end{eqnarray}
where
\begin{eqnarray}
\label{cdelta} c_\delta \!\!& \equiv &\!\!
\left[6 \! \int_{A_{\rm dec}}^1 \frac{dx}{x^4} \, \frac{E(x)}{H(x)/H_0} \right]^{\!-1} \!,
\\
\label{cSigma0} c_\Sigma \!\!& \equiv &\!\! E_0 c_\delta.
\end{eqnarray}
Here, $A_{\rm dec} = A(t_{\rm dec})$ is the mean expansion parameter evaluated at the
time of decoupling. In the following we simply assume that $A_{\rm dec} = 1/(1+z_{\rm dec})$,
where $z_{\rm dec} \simeq 1090$~\cite{WMAP7b} is the redshift at decoupling.

In Fig.~2, we plot the skewness and the actual shear as a function of
the mean dark energy density, for various values of the mean dark energy equation
of state parameter. It is evident from the figure that, for a wide range of
values of $\Omega_{\rm DE}$ and $w$, the parameters $\delta$ and $\Sigma_0$ are
indeed small quantities, as we have previously assumed.


\begin{figure}
\begin{center}
\includegraphics[clip,width=0.7\textwidth]{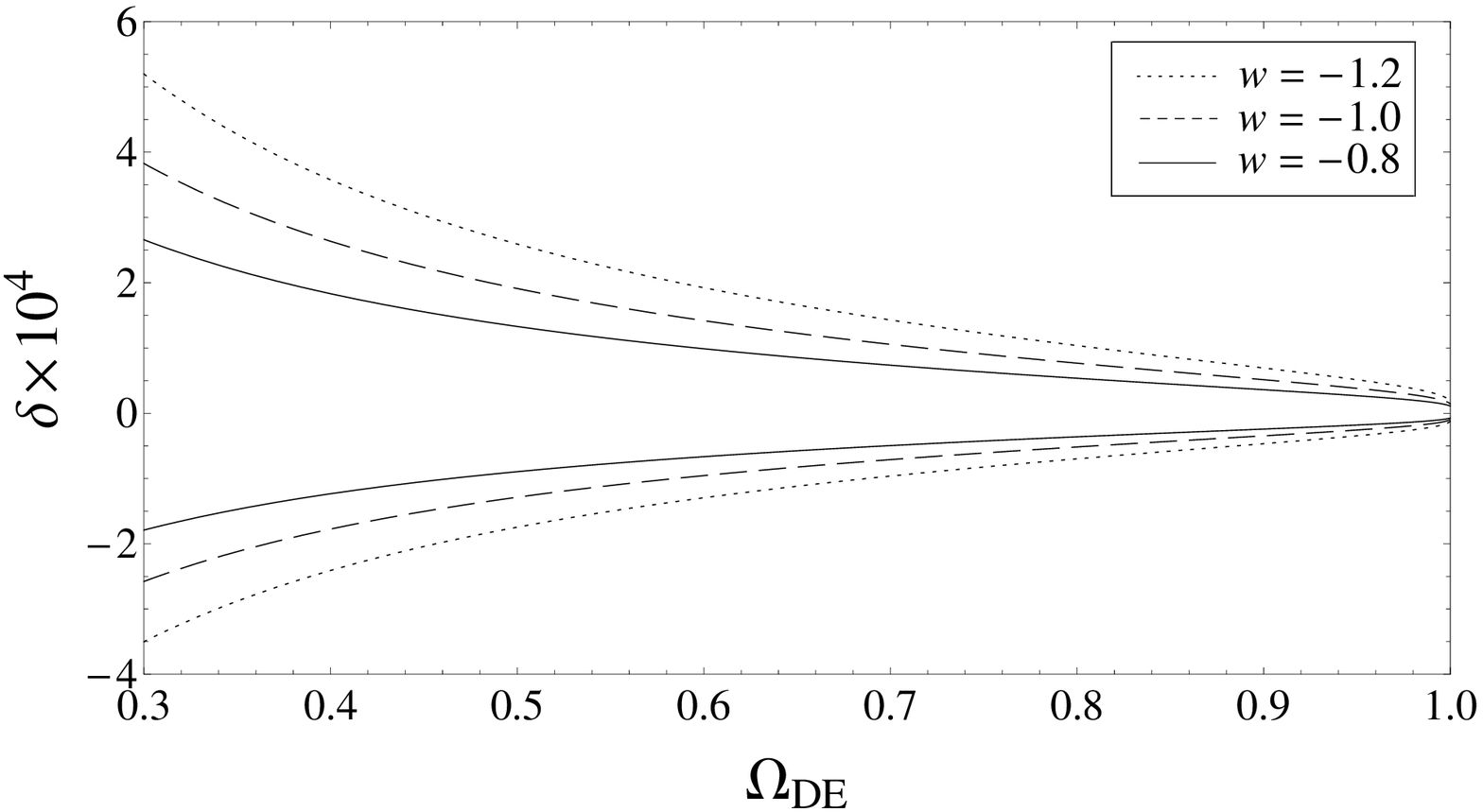}

\vspace{0.3cm}

\includegraphics[clip,width=0.7\textwidth]{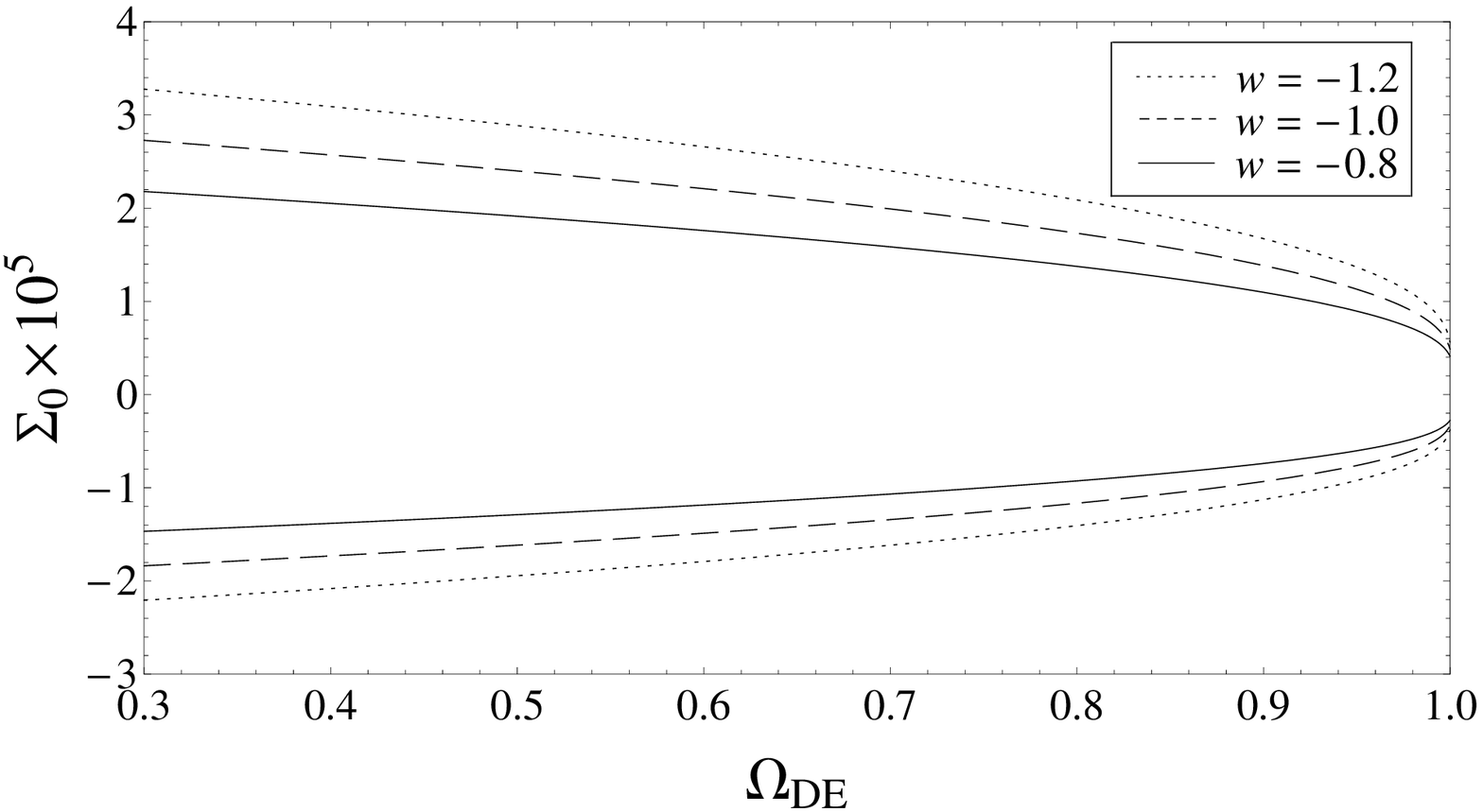}
\caption{The skewness $\delta$ (upper panel) and the shear $\Sigma_0$ (lower panel)
as a function of the mean dark energy density $\Omega_{\rm DE}$ for different values of the
mean dark energy equation of state parameter $w$. The values of $\delta$ and $\Sigma_0$
ensure that the eccentricity at decoupling is just that given by Eqs.~(\ref{QIWMAP})
and (\ref{ecc-dec}).}
\end{center}
\end{figure}


The smallness of the parameters of anisotropy, $\delta$ and $\Sigma_0$,
is a reason to believe that the values of mean parameters $\Omega_{\rm DE}$ and $w$
are very close to the analogous ones for the isotropic standard cosmological model,
which according the 7-yr WMAP data are~\cite{WMAP7b}:
\begin{eqnarray}
\label{OmegaStat} && \Omega_{\rm DE}^{(\rm isotropic)} = 0.734 \pm 0.029 \; (68\% \;\mbox{C.L.}), \\
\label{wStat}     && w^{(\rm isotropic)} = -0.980 \pm 0.053 \; (68\% \;\mbox{C.L.}).
\end{eqnarray}
From the above equations and looking at Fig.~2, we finally get
an order of magnitude estimate of $\delta$ and $\Sigma_0$:
\begin{equation}
\label{deltaSigma0stima} |\delta| \sim 10^{-4}, \;\;\; |\Sigma_0| \sim 10^{-5}.
\end{equation}
The above value of $\delta$ is compatible with the constraint obtained by Koivisto and
Mota~\cite{Koivisto-Mota} coming from the analysis of luminosity distance-redshift relation
of type Ia supernovae.~\footnote{The dark energy equation of state parameter $w$
and the skewness $\delta$ introduced in this paper correspond, respectively,
to the parameters $w +\gamma$ and $-3\gamma$ defined in Ref.~\cite{Koivisto-Mota}.
Translating the results of Ref.~\cite{Koivisto-Mota} to our case,
we get $-0.15 \leq \delta \leq 0.21 \; (68.3\% \;\mbox{C.L.})$.}
Instead, no constraint exists on $\Sigma_0$ in the literature.

Nevertheless, a full analysis of magnitude-redshift data on type Ia supernovae
can put limits both on $\delta$ and $\Sigma_0$ and, indeed, preliminary
results~\cite{InProgress} indicate that
\begin{eqnarray}
\label{OmegaStat} && -0.40 \leq \delta \leq 0.13 \;\; (1\sigma \; \mbox{C.L.}), \\
\label{wStat}     && -0.026 \leq \Sigma_0 \leq 0.014 \;\; (1\sigma \; \mbox{C.L.}).
\end{eqnarray}
This leaves open the possibility of having an ellipsoidal universe
{\it via} an anisotropic dark energy.


\section{\normalsize{IV. Conclusions}}
\renewcommand{\thesection}{\arabic{section}}

The 7-year WMAP observations~\cite{WMAP7a} do not have alleviated the so-called
quadrupole problem of CMB anisotropy spectrum, and a tension between data and the
quadrupole amplitude predicted by the best-fit $\Lambda$CDM concordance model still remains.

On the other hand, an anisotropic cosmological model described by a Bianchi type I metric,
the ``ellipsoidal universe''~\cite{prl,prd}, allows a lower quadrupole and
better matches the large-scale CMB anisotropy data.

In the seminal papers~\cite{Koivisto-Mota,Koivisto-Mota2} by Koivisto and Mota,
the peculiar features of a universe filled with an anisotropic dark energy fluid were
deeply and exhaustively studied.

In this paper, instead, we have investigated the possibility that a dark energy component
with anisotropic equation of state could generate an ellipsoidal universe with
the right characteristics to explain the low quadrupole in the CMB fluctuations.

The amount of anisotropy in the equation of state of dark energy,
the skewness $\delta$, is transferred to the background geometry which gets anisotropized
at a level that today is given the actual shear $\Sigma_0$.
Imposing that such a level of cosmic anisotropy is exactly that necessary to
generate a quadrupole term in the CMB spectrum in such a way to match
the ``low'' value of the observed one, we have obtained the estimates:
$|\delta| \sim 10^{-4}$ and $|\Sigma_0| \sim 10^{-5}$.

These values are well within the $1\sigma$ confidence region allowed by
magnitude-redshift data of type Ia supernovae~\cite{Koivisto-Mota,InProgress},
whose analysis constitutes, up to today, the only test available
in the literature constraining cosmic anisotropy of Bianchi type I.

This kind of test when combined with other tests on cosmic anisotropy,
such as that coming from the study of CMB polarization spectrum,
could put severe limits on the existence of anisotropic dark energy, confirm the
ellipsoidal universe proposal or rules it out. All of this is, however,
beyond the aim of this paper and will be the subject of future investigations.


\section{\normalsize{Appendix}}

The total CMB quadrupole intensity, $\mathcal{Q}_{\rm tot}$,
is given by Eqs.~(\ref{Cl})-(\ref{alm}) as~\cite{prl,prd}:
\begin{equation}
\label{Aquadrupole}
\mathcal{Q}_{\rm tot}^2 = \mathcal{Q}_{\rm A}^2 + \mathcal{Q}_{\rm I}^2 - 2f
\mathcal{Q}_{\rm A} \mathcal{Q}_{\rm I} \, ,
\end{equation}
where $\mathcal{Q}_{\rm I}$ and $\mathcal{Q}_{\rm A}$ are respectively given by Eqs.~(\ref{QIWMAP}) and (\ref{QA}),
\begin{equation}
\label{Af}
f(\vartheta, \varphi \, \phi_2, \phi_3) =  
             \frac{1 + 3 \cos(2 \vartheta) + 2 \sqrt{6} \sin\vartheta \left[\sin\vartheta \cos(2\varphi + \phi_3) -
             2 \cos\vartheta \cos(\varphi + \phi_2) \right] }{4\sqrt{5}} \, ,
\end{equation}
and we are considering, for the sake of simplicity, the case $a > b$.
In Eq.~(\ref{Af}), $\vartheta \in [0,\pi]$ and $\varphi \in [0,2\pi]$ are the polar angles defining,
in the galactic coordinate system, the direction of the axis of symmetry associated to planar symmetry of
ellipsoidal universe model~\cite{prl,prd}. Together with the
eccentricity at decoupling, $e_{\rm dec}$, they completely define the coefficients $a^{\rm A}_{2m}$.
The parameters $\phi_{2,3} \in [0,2\pi]$, instead, are unknown phases (which, roughly speaking, define the
``direction'' of the inflation-produced quadrupole)
that together with $\mathcal{Q}_{\rm I}$ completely define the coefficients $a^{\rm I}_{2m}$.
Explicitly, we have~\cite{prl,prd}:
\begin{eqnarray}
\label{AalmA}
&& a_{20}^{\rm A} = -\frac{\sqrt{\pi}}{6\sqrt{5}} \,
                  [1 + 3\cos(2 \vartheta) ] \, e_{\rm dec}^2 \, , \nonumber \\
&& a_{21}^{\rm A} = -(a_{2,-1}^{\rm A})^{*} =
                  \sqrt{\frac{\pi}{30}} \;
                  e^{-i \varphi}  \sin(2\vartheta) \, e_{\rm dec}^2 \, , \\
&& a_{22}^{\rm A} = (a_{2,-2}^{\rm A})^{*} = -
                  \sqrt{\frac{\pi}{30}}
                  \; e^{-2 i \varphi} \sin^2\!\vartheta \,
                  e_{\rm dec}^2 \, , \nonumber
\end{eqnarray}
and
\begin{eqnarray}
\label{AalmI}
&& a^{\rm I}_{20} = \sqrt{\frac{\pi}{3}} \; \mathcal{Q}_{\rm I} \, , \nonumber \\
&& a^{\rm I}_{21} =  - (a^{\rm I}_{2,-1})^{*} = \sqrt{\frac{\pi}{3}} \; e^{i \phi_2} \mathcal{Q}_{\rm I} \, , \\
&& a^{\rm I}_{22} =  (a^{\rm I}_{2,-2})^{*} = \sqrt{\frac{\pi}{3}}
\; e^{i \phi_3} \mathcal{Q}_{\rm I} \, . \nonumber
\end{eqnarray}
Let us assume that there exists some mechanism able to produce an anisotropization of the Universe,
so that $e_{\rm dec}$ is a parameter fixed by the model. Generally, on the other hand, the
direction of the symmetry axis $(\vartheta,\varphi)$ cannot be fixed by the model and should be
considered as unknown. The same its true for the direction of the inflation-produced quadrupole ($\phi_2,\phi_3$).
We therefore assume that $\vartheta$, $\varphi$, $\phi_2$, and $\phi_3$ are stochastic variables which assume
random values in their intervals of existence. In Fig.~3 (see the upper panel), we show the probability distribution function
of $f$, $F[f]$, derived by performing a Monte Carlo simulation.


\begin{figure}[t!]
\begin{center}
\includegraphics[clip,width=0.7\textwidth]{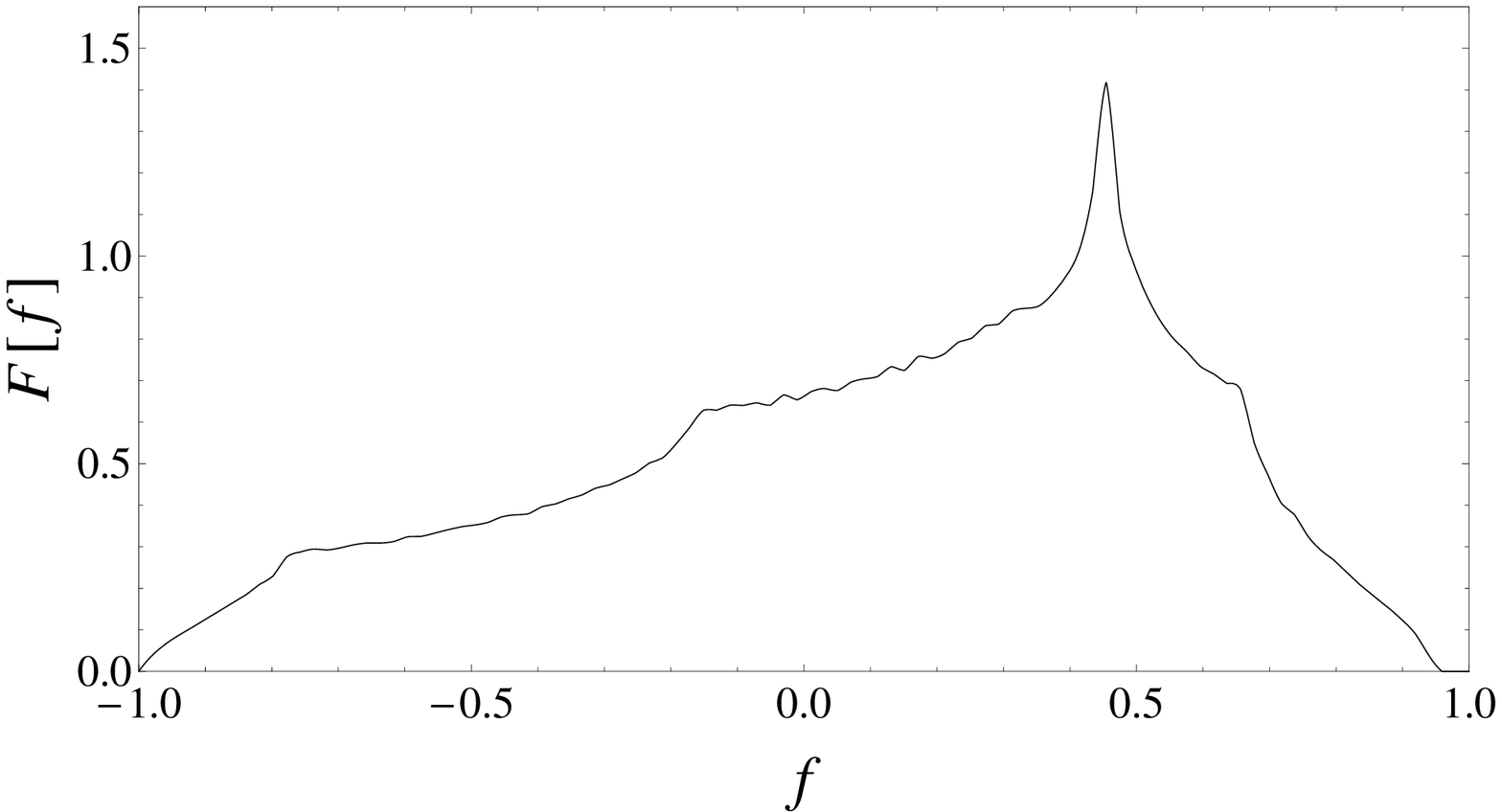}

\vspace{0.3cm}

\includegraphics[clip,width=0.7\textwidth]{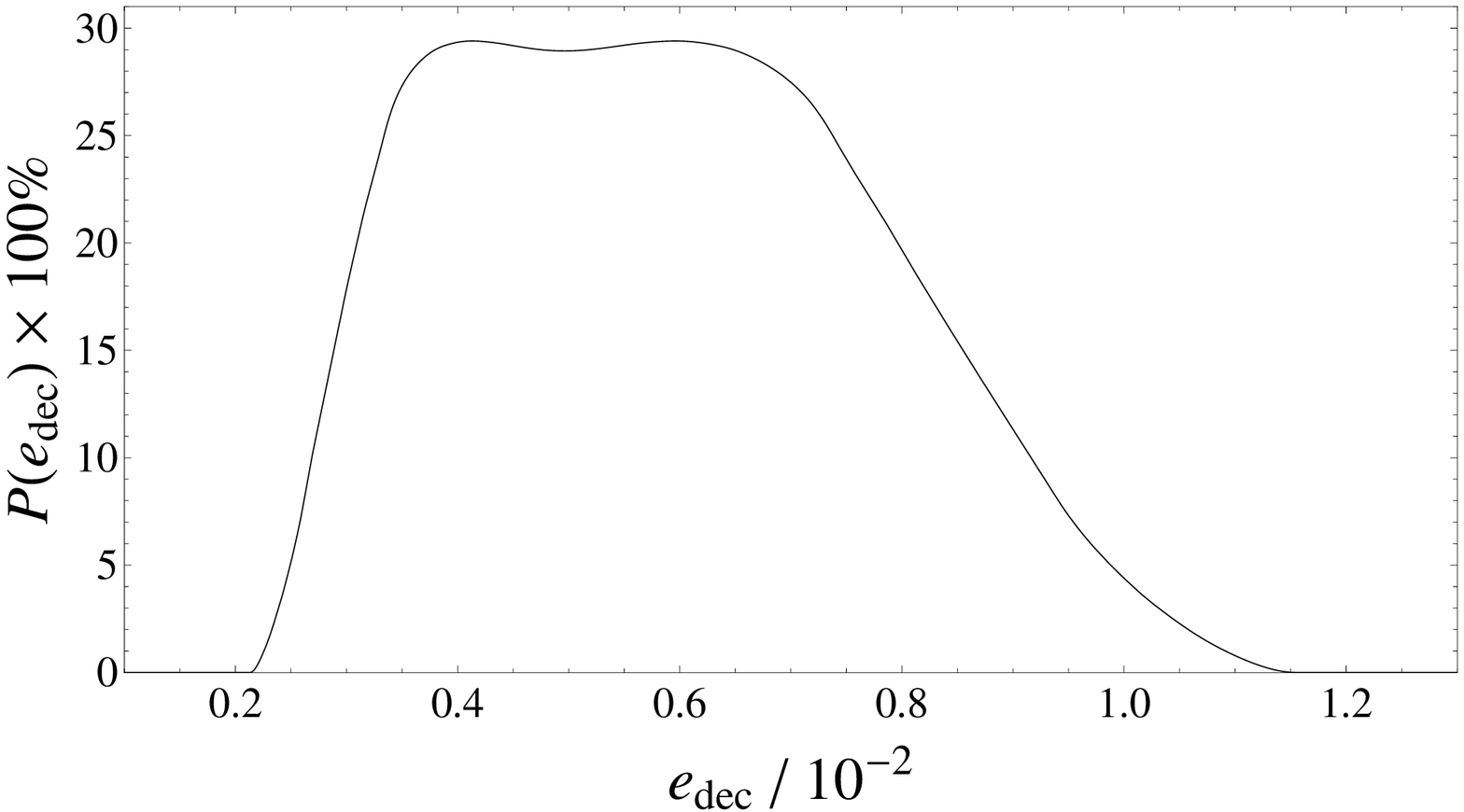}
\caption{The probability distribution function of $f$
(upper panel), and the probability
that the observed quadrupole $\mathcal{Q}^2_{\rm obs}$ is in the 1$\sigma$ confidence interval of
the theoretical total quadrupole $\mathcal{Q}_{\rm tot}^2$, at the varying of
the eccentricity at decoupling $e_{\rm dec}$ (lower panel).}
\end{center}
\end{figure}


We want now to calculate the probability, whatever are the values of $\vartheta$, $\varphi$, $\phi_2$, and $\phi_3$,
with $e_{\rm dec}$ being fixed, that the observed quadrupole $\mathcal{Q}^2_{\rm obs}$,
given in Eq.~(\ref{QWMAP}), is in the 1$\sigma$ confidence interval of $\mathcal{Q}_{\rm tot}^2$, namely
\begin{equation}
\label{AP}
P(e_{\rm dec}) \equiv P(e_{\rm dec} \, | \: \mathcal{Q}^2_{\rm obs} \!
\in [\mathcal{Q}_{\rm tot}^2 - \sigma_{\rm tot} \, , \, \mathcal{Q}_{\rm tot}^2 + \sigma_{\rm tot}]).
\end{equation}
The uncertainty $\sigma_{\rm tot}$ on $\mathcal{Q}^2$ is obtained by simply propagating
the uncertainty on the inflation-produced quadrupole $\mathcal{Q}_{\rm I}$, i.e. the cosmic
variance in Eq.~(\ref{CosmicVariance}), by means of Eq.~(\ref{Aquadrupole}):
\begin{equation}
\label{Asigma}
\sigma_{\rm tot} = \left| 1- f \, \frac{\mathcal{Q}_{\rm A}}{\mathcal{Q}_{\rm I}} \right| \sigma_{\rm cosmic} \, .
\end{equation}
It is easy to see that the condition
$\mathcal{Q}^2_{\rm obs} \! \in [\mathcal{Q}_{\rm tot}^2 - \sigma_{\rm tot} \, , \, \mathcal{Q}_{\rm tot}^2 + \sigma_{\rm tot}]$
is equivalent to the condition $f \in [f_{-}, f_{+}]$, where
\begin{equation}
\label{Afpm}
f_{\pm}(e_{\rm dec}) = \frac{1 \pm \sqrt{\frac25} + \mathcal{Q}_{\rm A}^2 - \mathcal{Q}_{\rm obs}^2}
{\left(2 \mp \sqrt{\frac25}\right) \! \mathcal{Q}_{\rm A} \mathcal{Q}_{\rm I}} \, .
\end{equation}
We then have that
\begin{equation}
\label{AProb}
P(e_{\rm dec}) = \int_{f_{-}}^{f_{+}} \! df F[f]
\end{equation}
gives the probability, at fixed $e_{\rm dec}$, that
a cancelation between the quadrupole associated to the
anisotropic spacetime background and the inflation-produced one, so to explain the ``quadrupole problem'',
occurs coincidentally (i.e., whatever is the direction of the axis of symmetry).
The lower panel of Fig.~3 shows $P(e_{\rm dec})$ as a function of $e_{\rm dec}$,
and indicates that such a probability is not negligible for $e_{\rm dec}$ in the range
$[0.25,1.15] \times 10^{-2}$. In particular, $P(e_{\rm dec}) \simeq 29\%$ for
$e_{\rm dec} \simeq 0.65 \times 10^{-2}$ given by Eq.~(\ref{ecc-dec}), which amounts to have
$\mathcal{Q}_{\rm tot} \simeq \mathcal{Q}_{\rm obs}$~\cite{prl,prd}.



\begin{thebibliography}{99}

\bibitem{COBE1}            G.~F.~Smoot {\it et al.},
                           Astrophys.\ J.\ {\bf 396}, L1 (1992).

\bibitem{WMAP1}            G.~Hinshaw {\it et al.}  [WMAP Collaboration],
                           Astrophys.\ J.\ Suppl.\ {\bf 148}, 135 (2003).

\bibitem{COBE2}             C.~L.~Bennett {\it et al.},
                           Astrophys.\ J.\ {\bf 464}, L1 (1996).

\bibitem{Weinberg}         S. Weinberg,
                           {\it Cosmology}
                           (Oxford University Press, New York, New York, 2008).

\bibitem{WMAP7a}           D.~Larson {\it et al.},
                           arXiv:1001.4635 [astro-ph.CO].

\bibitem{PLANCK}           PLANCK mission website:\\
                           http://www.rssd.esa.int/index.php?project=PLANCK

\bibitem{Liu}              H.~Liu and T.~P.~Li,
                           arXiv:0911.4063 [astro-ph.CO];
                           arXiv:1003.1073 [astro-ph.CO].

\bibitem{Quadrupole}       G.~Efstathiou,
                           Mon.\ Not.\ Roy.\ Astron.\ Soc.\ {\bf 343}, L95 (2003);
                           B.~Feng and X.~Zhang,
                           Phys.\ Lett.\ B {\bf 570}, 145 (2003);
                           M.~Kawasaki and F.~Takahashi,
                           Phys.\ Lett.\ B {\bf 570}, 151 (2003);
                           J.~M.~Cline, P.~Crotty and J.~Lesgourgues,
                           JCAP {\bf 0309}, 010 (2003);
                           C.~Gordon and W.~Hu,
                           Phys.\ Rev.\ D {\bf 70}, 083003 (2004);
                           T.~Moroi and T.~Takahashi,
                           Phys.\ Rev.\ Lett.\ {\bf 92}, 091301 (2004);
                           Y.~S.~Piao,
                           Phys.\ Rev.\ D {\bf 71}, 087301 (2005);
                           D.~Boyanovsky, H.~J.~de Vega and N.~G.~Sanchez,
                           Phys.\ Rev.\ D {\bf 74}, 123006 (2006);
                           Phys.\ Rev.\ D {\bf 74}, 123007 (2006);
                           P.~Cea,
                           arXiv:astro-ph/0702293;
                           M.~Demianski and A.~G.~Doroshkevich,
                           Phys.\ Rev.\ D {\bf 75}, 123517 (2007);
                           A.~Rakic and D.~J.~Schwarz,
                           Phys.\ Rev.\ D {\bf 75}, 103002 (2007);
                           X.~H.~Ge and S.~P.~Kim,
                           JCAP {\bf 0707}, 001 (2007);
                           A.~Gruppuso,
                           Phys.\ Rev.\ D {\bf 76}, 083010 (2007);
                           C.~G.~Boehmer and D.~F.~Mota,
                           Phys.\ Lett.\ B {\bf 663}, 168 (2008);
                           C.~Destri, H.~J.~de Vega and N.~G.~Sanchez,
                           Phys.\ Rev.\ D {\bf 78}, 023013 (2008);
                           S.~Lee,
                           arXiv:0811.1643 [gr-qc];
                           A.~Pontzen,
                           Phys.\ Rev.\ D {\bf 79}, 103518 (2009);
                           L.~P.~He and Q.~Guo,
                           Res.\ Astron.\ Astrophys.\ {\bf 10}, 116 (2010)
                           [arXiv:0912.1913 [astro-ph.CO]];
                           L.~Grisa and L.~Sorbo,
                           arXiv:1002.0510 [astro-ph.CO];
                           D.~Caceres, L.~Castaneda and J.~M.~Tejeiro,
                           arXiv:1003.3491 [gr-qc];
                           L.~R.~Abramo and T.~S.~Pereira,
                           arXiv:1002.3173 [astro-ph.CO];
                           P.~Cea,
                           Mon.\ Not.\ Roy.\ Astron.\ Soc.\ {\bf 406}, 586 (2010)
                           [arXiv:1001.2650 [astro-ph.CO]].

\bibitem{prl}              L.~Campanelli, P.~Cea and L.~Tedesco,
                           Phys.\ Rev.\ Lett.\ {\bf 97}, 131302 (2006)
                           [Erratum-ibid.\ {\bf 97}, 209903 (2006)].

\bibitem{prd}              L.~Campanelli, P.~Cea and L.~Tedesco,
                           Phys.\ Rev.\ D {\bf 76}, 063007 (2007).

\bibitem{Berera}           A.~Berera, R.~V.~Buniy and T.~W.~Kephart,
                           JCAP {\bf 0410}, 016 (2004).

\bibitem{prd2}             L.~Campanelli,
                           Phys.\ Rev.\ D {\bf 80}, 063006 (2009).

\bibitem{Beltran Jimenez}  J.~Beltran Jimenez and A.~L.~Maroto,
                           Phys.\ Rev.\ D {\bf 76}, 023003 (2007).

\bibitem{Koivisto-Mota}    T.~Koivisto and D.~F.~Mota,
                           Astrophys.\ J.\ {\bf 679}, 1 (2008).

\bibitem{Koivisto-Mota2}   T.~Koivisto and D.~F.~Mota,
                           JCAP {\bf 0806}, 018 (2008).

\bibitem{Barrow}           J.~D.~Barrow,
                           Phys.\ Rev.\ D {\bf 55}, 7451 (1997).

\bibitem{Rodrigues}        D.~C.~Rodrigues,
                           Phys.\ Rev.\ D {\bf 77}, 023534 (2008).

\bibitem{Akarsu}           O.~Akarsu and C.~B.~Kilinc,
                           arXiv:0807.4867 [gr-qc].

\bibitem{Appleby}          S.~Appleby, R.~Battye and A.~Moss,
                           arXiv:0912.0397 [astro-ph.CO].

\bibitem{Battye}           R.~Battye and A.~Moss,
                           Phys.\ Rev.\ D {\bf 80}, 023531 (2009).

\bibitem{Quercellini}      C.~Quercellini, M.~Quartin and L.~Amendola,
                           Phys.\ Rev.\ Lett.\ {\bf 102}, 151302 (2009);
                           C.~Quercellini, P.~Cabella, L.~Amendola, M.~Quartin and A.~Balbi,
                           Phys.\ Rev.\ D {\bf 80}, 063527 (2009);
                           C.~Quercellini, L.~Amendola, A.~Balbi, P.~Cabella and M.~Quartin,
                           arXiv:1011.2646 [astro-ph.CO].

\bibitem{Taub}             A.~H.~Taub,
                           Annals Math.\ {\bf 53}, 472 (1951).

\bibitem{Bunn}             E.~F.~Bunn and A.~Bourdon,
                           Phys.\ Rev.\ D {\bf 78}, 123509 (2008).

\bibitem{Durrer}           R.~Durrer,
                           {\it The Cosmic Microwave Background}
                           (Cambridge University Press, Cambridge, United Kingdom, 2008).

\bibitem{Kolb}             E.~W.~Kolb and M.~S.~Turner,
                           {\it The Early Universe}
                           (Addison-Wesley, Redwood City, California, 1990).

\bibitem{WMAP7b}           E.~Komatsu {\it et al.},
                           arXiv:1001.4538 [astro-ph.CO].

\bibitem{InProgress}       L.~Campanelli, P.~Cea, G.~L.~Fogli and A.~Marrone,
                           arXiv:1012.5596 [astro-ph.CO].

\end{thebibliography}
\end{document}